\documentclass[a4paper,fleqn,usenatbib]{mnras}

\usepackage{newtxtext,newtxmath}
\usepackage[flushleft]{threeparttable}
\usepackage{times,epsfig,natbib,graphics,longtable,lscape}
\usepackage[edges]{forest}
\usepackage{pdflscape}

\usepackage[T1]{fontenc}
\usepackage{ae,aecompl}

\usepackage{nccmath}
\usepackage{multicol}

\defcitealias{LVCGCNS190425z}{24168}
\defcitealias{LVCGCNS190426c}{24237}
\defcitealias{LVCGCNS190901ap}{25606}
\defcitealias{LVCGCNS190910d}{25695}
\defcitealias{LVCGCNS190923y}{25814}
\defcitealias{LVCGCNS191205ah}{26350}
\defcitealias{LVCGCNS191213g}{26402}
\defcitealias{LVCGCNS200115j}{26759}
\defcitealias{LVCGCNS200213t}{27042}

\title[Searching for electromagnetic counterparts of gravitational-wave events with ASAS-SN]{ASAS-SN search for optical counterparts of gravitational-wave events from the third observing run of Advanced LIGO/Virgo}

\author[de Jaeger et al.]
{T. de Jaeger$^{1}$\thanks{E-mail: dejaeger@hawaii.edu},
B. J. Shappee$^{1}$,
C. S. Kochanek$^{2,3}$,
K. Z. Stanek$^{2,3}$,
J.~F.~Beacom$^{4,2,3}$,
\newauthor
T.~W.-S.~Holoien$^{5}$\thanks{NHFP Einstein Fellow},
Todd A. Thompson$^{2,3}$,
A. Franckowiak$^{6}$,
S. Holmbo$^{7}$
\\
\small
$^{1}$ Institute for Astronomy, University of Hawaii, 2680 Woodlawn Drive, Honolulu, HI 96822, USA.\\
$^{2}$ Department of Astronomy, The Ohio State University, 140 W. 18th Avenue, Columbus, OH 43210, USA.\\
$^{3}$ Center for Cosmology and AstroParticle Physics (CCAPP), The Ohio State University, 191 W. Woodruff Avenue, Columbus, OH 43210, USA.\\
$^{4}$ Department of Physics, The Ohio State University, 191 W. Woodruff Ave., Columbus, OH 43210, USA.\\
$^{5}$ The Observatories of the Carnegie Institution for Science, 813 Santa Barbara St., Pasadena, CA 91101, USA.\\
$^{6}$ Fakult{\"a}t f{\"u}r Physik \& Astronomie, Ruhr-Universit{\"a}t Bochum, D-44780 Bochum, Germany.\\
$^{7}$Department of Physics and Astronomy, Aarhus University, Ny Munkegade 120, DK-8000 Aarhus C, Denmark.\\}

\date{}

\pubyear{2021}

\begin{document}
\label{firstpage}
\pagerange{\pageref{firstpage}--\pageref{lastpage}}

\maketitle

\begin{abstract}
\noindent
We report on the search for electromagnetic counterparts to the nine gravitational-wave events with a $>$60\% probability of containing a neutron star during the third (O3) LIGO-Virgo Collaboration (LVC) observing run with the All-Sky Automated Survey for SuperNovae (ASAS-SN). No optical counterparts associated with a gravitational wave event was found. However, thanks to its network of telescopes, the average area visible to at least one ASAS-SN site during the first 10 hours after the trigger contained $\sim$30\% of the integrated source location probability. Through a combination of normal operations and target-of-opportunity observations, ASAS-SN observations of the highest probability fields began within one hour of the trigger for four of the events. After 24 hours, ASAS-SN observed $>$60\% of total probability for three events and $>$40\% for all but one of the events. This is the largest area coverage to a depth of $g = 18.5$ mag from any survey with published coverage statistics for seven of the nine events. With its observing strategy, five sites around the world, and a large field of view, ASAS-SN will be one of the leading surveys to optically search for nearby neutron star mergers during LVC O4.

\end{abstract}



\section{Introduction}

Since the beginning of astronomy, we have used electromagnetic (EM) radiation to explore and understand our Universe. Gravitational waves (GW) opened a new window for astronomy with the first-ever direct observation of GW on September 14, 2015 by two detectors of the Laser Interferometer Gravitational-Wave Observatory (LIGO; \citealt{aasi15,LIGO2016}). Since this first GW detection, $\sim$70 events have been observed, with a majority due to merging black holes (BH-BH; \citealt{abbott_O1O2,abbottO3}). In classical general relativity, BH-BH mergers are generally not expected to have EM counterparts \citep{abbott16a}. No confirmed counterparts have been detected for any of BH-BH mergers seen to date (but see \citealt{graham20} for a possibility), although several theoretical studies have suggested possible EM emission mechanisms (e.g., \citealt{palenzuela10,moesta12}).

Unlike BH-BH mergers, binary neutron star merger (BNS) mechanisms are expected to yield an optical counterparts powered by the radioactive decay of rapid neutron capture process (r-process) elements synthesised in the merger ejecta \citep{li98,metzger10} or by the cooling of shock-heated material around the neutron star \citep{piro18}. On August 17, 2017, the LVC detected the first and best example of a BNS merger: GW170817 \citep{gw170817}. Only two seconds after its detection, a short gamma-ray burst (GRB170817A) was detected \citep{connaughton17,goldstein17}. GRB170817A was followed by the discovery of the optical counterpart SSS17a (AT2017gfo) $\sim$11 hours later in the galaxy NGC 4993 \citep{coulter17} and later confirmed by others teams \citep{smartt17,arcavi17,soares17,valenti17}. 

The combination of EM and GW information on the BNS can be used to constrain the mass and the radii of NS \citep{margalit17} or their equation of state \citep{bausewein12,annala18}. Also, observations of the optical-IR ``kilonova'' counterpart \citep{chornock17,cowperthwaite17,drout17,kilpatrick17,nicholl17,shappee17,valenti17,villar17}, provided the first observational confirmation that NS mergers produce the majority of the r-process elements heavier than iron \citep{burbidge75,cameron75,kasen17,pian17,metzger19}. They also permit tests of theoretical kilonova models. For example, \citet{drout17} showed that the temperatures cooled from 10,000 K to 5,100 K in between 12 and 36 hours after the event, confirming model predictions. Also, \citet{shappee17} used spectra taken 11.76 and 12.72 hours after the merger to show that the photosphere was expanding at $\sim$0.3 c.
 
Mergers with optical counterparts can also be used as standard sirens, allowing source distances
to be measured directly \citep{schutz86}. From only a single EM GW counterpart detection (GW170817), \citet{abbott17} derived a Hubble constant of $70^{+12}_{-8}$\,km\,s$^{-1}$\,Mpc$^{-1}$. Finally, any arrival time differences between the EM and GW signals test for propagation velocity differences that probe the nature of dark energy and can rule out theoretical models \citep{ezquiaga17,sakstein17,rubin20}. 

Advanced LIGO \citep{aasi15} and Advanced Virgo \citep{acernese15} started their third observing run (O3) to search for GW sources in April 2019. Even though the hunt was stopped a month early due to the COVID-19 pandemic, an increase in sensitivity during O3 \citep{ligorunO3} allowed the discovery of more than 56 events in 330 days, including nine with a high probability of having at least one NS. Unlike O1 and O2, LVC O3 had a public alert system that automatically sent alerts within minutes of the GW detections, allowing all astronomers to search for EM counterparts. Following the successful observations of SSS17a (GW170817), many surveys participated in the searches for GW optical counterparts during the LVC O3 run. These surveys include: the All-Sky Automated Survey for Supernovae (ASAS-SN; \citealt{Shappee14}), the Asteroid Terrestrial-impact Last Alert System (ATLAS; \citealt{Tonry18}), the Dark Energy Camera (DECam; \citealt{flaugher15}), the Electromagnetic counterparts of gravitational wave sources at the Very Large Telescope (ENGRAVE; \citealt{levan20}), the Gravitational-wave Optical Transient Observer (GOTO; \citealt{dyer18}), the Global Rapid Advanced Network Devoted to the Multi-messenger Addicts survey (GRANDMA; \citealt{antier20}), the Katzman Automatic Imaging Telescope (KAIT; \citealt{filippenko01}), the Panoramic Survey Telescope and Rapid Response System (Pan-STARRS; \citealt{chambers16}), the Searches After Gravitational-waves Using ARizona Observatories (SAGUARO; \citealt{lundquist19}), SkyMapper \citep{chang21}, and the Zwicky Transient Facility (ZTF; \citealt{bellm18}).

During the LVC O3, more than 1,500 Gamma-ray Coordinates Network circulars (GCN) were published to report EM counterpart searches and potential candidates. Approximately sixty observatories were involved in the optical follow-ups, yielding $\sim$388 candidate counterparts \citep{coughlin20b}. Unfortunately, no optical counterparts were linked to a GW event. This shows the uniqueness of SSS17a (GW170817; \citealt{gw170817}): a near perfect combination of a small 28 square degrees high probability localisation region and a distance of only 40$^{+14}_{-8}$ Mpc. In comparison, the events with a $>60\%$ probability of containing a NS in LVC O3 had an average 90\% confidence localisation of $\sim$4,700 square degrees and an average distance of $\sim$330 Mpc (see Table \ref{tab:LVC_events}). This explains why searches for EM counterparts in O3 were challenging and unsuccessful.

Several groups participating in searches for GW optical counterparts during LVC O3 run have already published an overview of their strategy: GOTO \citep{gompertz20}, GRANDMA \citep{antier20}, SAGURO \citep{paterson21}, and ZTF \citep{coughlin20,coughlin20c,kasliwal20}. These studies describe both individual events and the lessons learned from LVC O3 to prepare for LVC O4 (starting in $\sim$June 2022). For example, one concern that should be addressed by the GW community is the lack of follow-up for a large number of candidates even as several candidates were classified multiple times by different groups \citep{coughlin20c}. 

In this paper, we present a summary of the ASAS-SN observational strategy during the LVC O3 run and show that, thanks to our five sites spread over the world, the majority of the events were observable from at least one ASAS-SN site soon after their discovery. In section \ref{sec:strategy}, we describe ASAS-SN characteristics and our observational strategy. In section \ref{txt:events}, we present all the LVC events for which ASAS-SN obtained data during the O3 run, and finally, we discuss our results in section \ref{txt:discussion} and conclude in Section \ref{txt:conclusions}.

\section{The ASAS-SN strategy}\label{sec:strategy}

ASAS-SN is the first and still the only ground-based survey to map the entire visible sky daily to a depth of $g = 18.5$ mag \citep{Shappee14, Kochanek17b}. Currently, ASAS-SN is composed of five stations: Brutus, the original ASAS-SN unit, located on Haleakala in Hawaii (USA); Cassius and Paczynski, both situated at Cerro Tololo International Observatory (CTIO) in Chile; Leavitt at McDonald Observatory in Texas (USA), and finally Payne-Gaposchkin at the South African Astrophysical Observatory (SAAO) in Sutherland, South Africa. All units are hosted by the Las Cumbres Observatory Global Telescope network (LCOGT; \citealt{Brown13}) and consist of four 14-cm aperture Nikon telephoto lenses, each with a 4.47 by 4.47-degree field-of-view. ASAS-SN survey operations are composed of three dithered 90 second exposures using a $g$-band filter (all five units) with $\sim$15 seconds of overheads between each image, for a total of 315 seconds per field. Thanks to the five stations, the instantaneous field of view is roughly 360 square degrees. Note that we will refer to the estimated location of a GW event as the search region. In practice, the search region is defined by probabilities, and unless otherwise specified, we are considering the region enclosing 90\% of the probability for the event location. We try to observe fields in order of their average probability of containing the event.

\begin{figure} 
  \centering 
	\includegraphics[width=1.0\columnwidth]{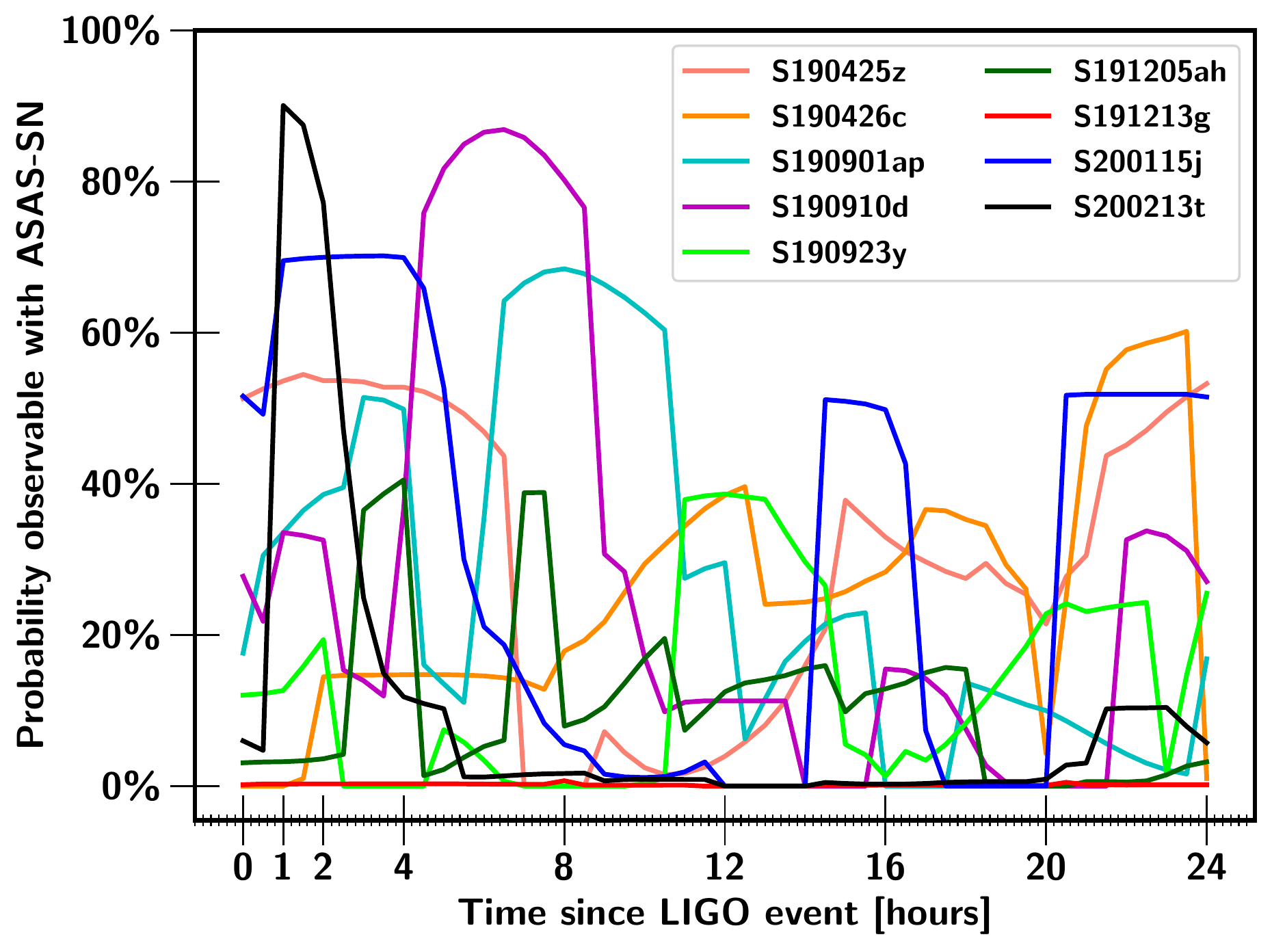}
  \caption{Instantaneous probability observable by ASAS-SN versus the time in hours since the LVC alert.}\label{fig:observability}
\end{figure}

In Figure \ref{fig:observability}, we show the fraction of the search region visible to ASAS-SN as a function of time for all the LVC O3 events with a $>$ 60\% probability that at least one of the compact objects is a NS. For all events, a large percentage of the search region ($\sim$30\%) is visible at least once within the first 24 hours. For example, less than one hour after the LVC alert, over 35\% of the search region was visible for five events while only three were under 15\% and one region was not visible at all. This allows ASAS-SN to respond promptly to almost all the LVC alerts independent of the time or localisation.

\begin{figure*} 
  \centering 
	\includegraphics[width=1.0\textwidth]{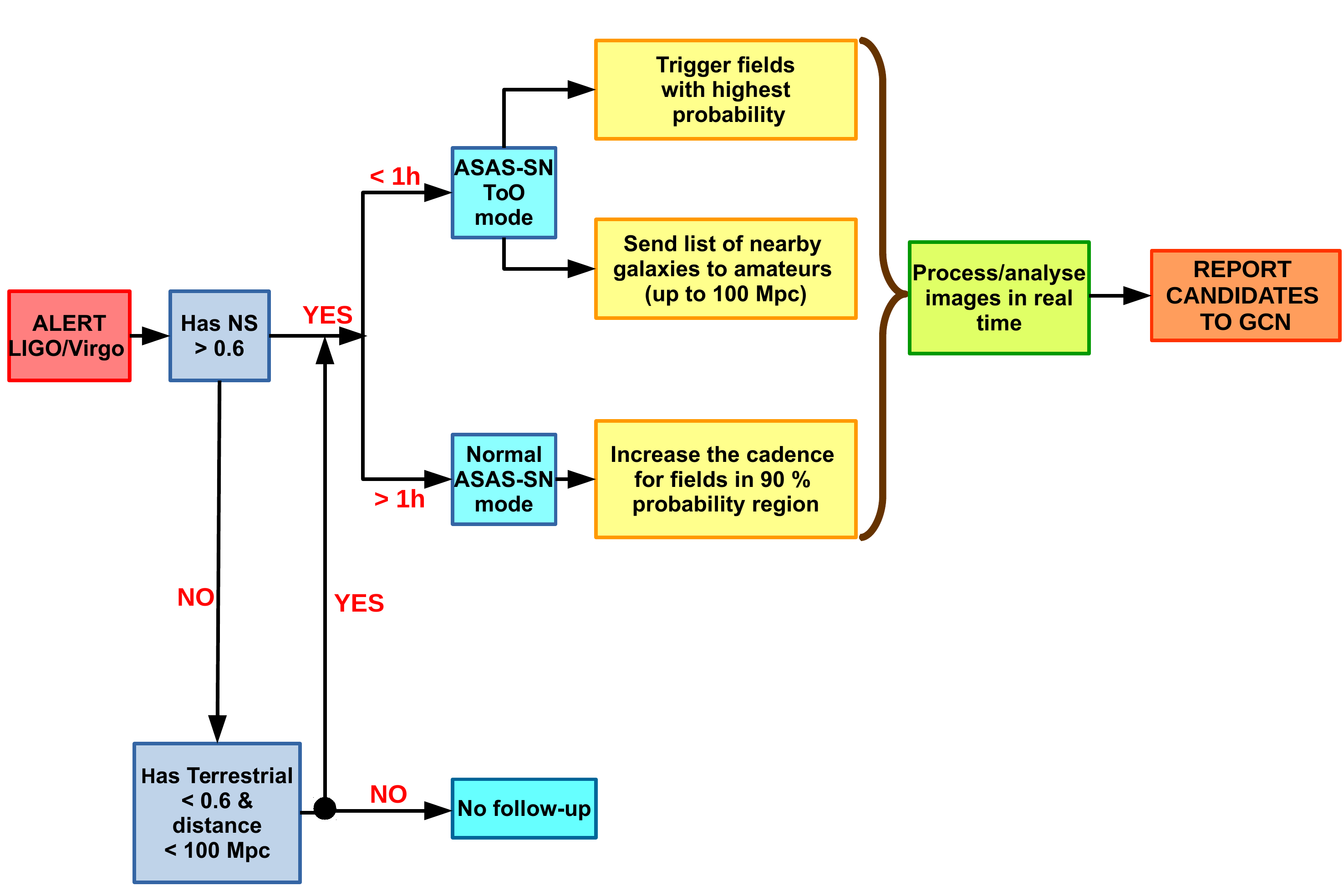}
  \caption{ASAS-SN strategy diagram. Our automated system receives a GCN LVC alert and updates the plan of observations to search for optical counterparts in only a few seconds. If the alert properties meet the triggering criteria, the list of fields/pointings to observe is automatically created and prioritised and all the images are analysed in real-time. Finally, good candidates are manually reported to GCN.}\label{fig:strategy}
\end{figure*}

To maximise our chance of observing an optical counterpart of GW sources, ASAS-SN has a Target-of-Opportunity mode (ToO) to enable rapid imaging follow-up. Our automated system receives all LVC alerts using the GCN protocol along with the IceCube real-time alerts \citep{aartsen17}. For all the GW events with either a (HasNS)$>60\%$ probability of containing a NS or a distance smaller than 100 Mpc and a $<40\%$ probability of being terrestrial, a list of the 30 pointings (each pointing corresponds to four cameras/fields, so 70 square degrees) with the highest localisation probability is automatically generated and the ToO observations start. After the first hour in ToO mode, the routine survey is optimised to increase the cadence of the fields in the search region. For example, during the LVC O3 run, pointings within the highest probability region were observed 4--5 times in 24 hours. All the images obtained from the ToO or the normal survey are processed and analysed in real-time, and the interesting optical counterpart candidates are reported to GCN. As described in Section \ref{txt:events}, no good candidates were detected and reported for the LVC O3. A schematic summary of the strategy is displayed in Figure \ref{fig:strategy}.
 
Finally, as a galaxy-targeted strategy is more suitable for telescopes with a small field-of-view, a list of nearby galaxies ($<$ 100 Mpc) is sent simultaneously to a group of dedicated amateur astronomers with whom ASAS-SN collaborates during normal operation to confirm ASAS-SN discoveries. Nearby galaxies are selected from the Gravitational Wave Galaxy Catalog \citep{white11}, a catalogue of 53,255 galaxies with high completeness to 100 Mpc and ranked by the probability that a particular galaxy hosts a GW event ($P_{\rm gal}$). The final probability is taken from \citet{coulter17} and depends on the search region ($P_{\rm 2D}$), the galaxy and event distances ($D_{\rm gal}$ and $D_{\rm LVC}$ respectively), the galaxy luminosity ($\tilde{L}_{B}$), and a normalisation factor ($k$):

\begin{equation}
  P_{\rm gal} = k^{-1} \times \tilde{L}_{B} \times P_{\rm 2D} \times \left ( 1 - \rm{erf} \left ( \frac{| D_{\rm gal} - D_{\rm LVC} |}{\sigma_{D,{\rm ~gal}}^{2} + \sigma_{D, {\rm ~LVC}}^{2}} \right ) \right )
\end{equation}
Given the completeness of the Gravitational Wave Galaxy Catalog and the large distance estimates for all of the LVC O3 events, it is unsurprising that the amateur astronomers did not find anything this cycle. They will continue to search in future observing runs.

\begin{table*}
\setlength\tabcolsep{2pt}
\begin{tabular}{llrlllrrrrrrl}

\hline\hline
 & & & & &  & \multicolumn{5}{c}{Classification Probability} & \\
 \cline{7-11}
Event &Date &90\% sky area & Distance & FAR &HasNS & BH-BH & NSBH & BNS & MassGap &Terrestrial &Prob. cov~  & GCN\\
 & UTC & deg$^{2}$ & Mpc &yr$^{-1}$ &[\%] &[\%] &[\%] &[\%] &[\%]  &[\%] &[\%] &\\
 \hline
S190425z  & April 25, 2019,
08:18:05 & 10183 & 155 (45) & $1.43\times 10^{-5}$ & >99 & <1 & <1 & >99 & <1 &0 & 67 & \citetalias{LVCGCNS190425z}\\
S190426c  & April 26, 2019,
15:21:55 & 1262 & 375 (108) & 0.61 &>99 & <1 & 13 & 49 & 24 &14 & 85 & \citetalias{LVCGCNS190426c}\\
S190901ap  & Sept. 1, 2019,
23:31:01 & 13613 & 242 (81) & 0.22 & >99 & <1 & <1 & 86 &<1  & 14 & 58 &  \citetalias{LVCGCNS190901ap}  \\
S190910d   & Sept. 10, 2019,
01:26:19 & 3829 & 606 (197) & 0.12 & >99 & <1 & 98 & <1   & <1 & 2 & 52 & \citetalias{LVCGCNS190910d}  \\
S190923y  & Sept. 23, 2019,
12:55:59 & 2107 & 438 (133) & 1.51 & >99 & <1 & 68 & <1 & <1 &32 & 44  &\citetalias{LVCGCNS190923y}  \\
S191205ah  &Dec. 5, 2019,
21:52:08  & 6378 & 385 (164) & 0.39 & >99 & <1 & 93 & <1 & <1  & 7 & 42&\citetalias{LVCGCNS191205ah}  \\
S191213g  &Dec. 13, 2019,
04:34:08 & 1393 & 195 (59) & 1.12 & >99 & <1 & <1 & 77 & <1 & 23 & 1 &\citetalias{LVCGCNS191213g}  \\
S200115j  & Jan. 15, 2020,
04:23:09  & 908 & 331 (97) & $6.61\times 10^{-4}$ & >99 & <1 & <1 & <1 &>99 & <1 & 63 &\citetalias{LVCGCNS200115j}  \\
S200213t  & Feb. 13, 2020,
04:10:40 & 2587 & 224 (90) & 0.56 & >99 & <1 & <1 & 63 & <1 & 37 & 42 &\citetalias{LVCGCNS200213t}  \\
\hline\hline
\end{tabular}
\caption{Characteristics of the LVC events triggered by ASAS-SN during the O3 run. In the first column the event name, followed by its discovery date in UTC are listed. In column 3 we list the sky area in square degrees of the search region. In columns 4 and 5 we list the distance in Mpc together with its uncertainty and their False Alarm Rate (FAR) in year$^{-1}$. In column 6, we list the probability that the lighter compact object has a mass $<$ 3 solar masses while in columns 7, 8, 9, 10, and 11 give the classification probabilities of the GW signal. Column 12 is the fraction of the probability that ASAS-SN covered in 24 hours. Finally, GCN circular reference is given in the last column. All the parameters were taken from the Gravitational-Wave Candidate Event Database (\url{https://gracedb.ligo.org/superevents/public/O3/}).}
\label{tab:LVC_events}
\end{table*}

\section{EW follow-up}\label{txt:events}

In this section, we describe each of the nine events announced for the LVC O3 run that passed our selection criterion as described in Section \ref{sec:strategy}. For the nine events, all having a (HasNS)$>60\%$ probability of containing a NS, we obtained data from a combination of normal operations and an optimisation of our observational strategy to observe the search region with a higher cadence. For all the events presented in this section, we use the initial localisation map from the BAYESian TriAngulation and Rapid localization (BAYESTAR; \citealt{singer16}). A list of those events together with their principal characteristics is given in Table \ref{tab:LVC_events}.

\begin{figure*}
	\includegraphics[width=1.0\textwidth]{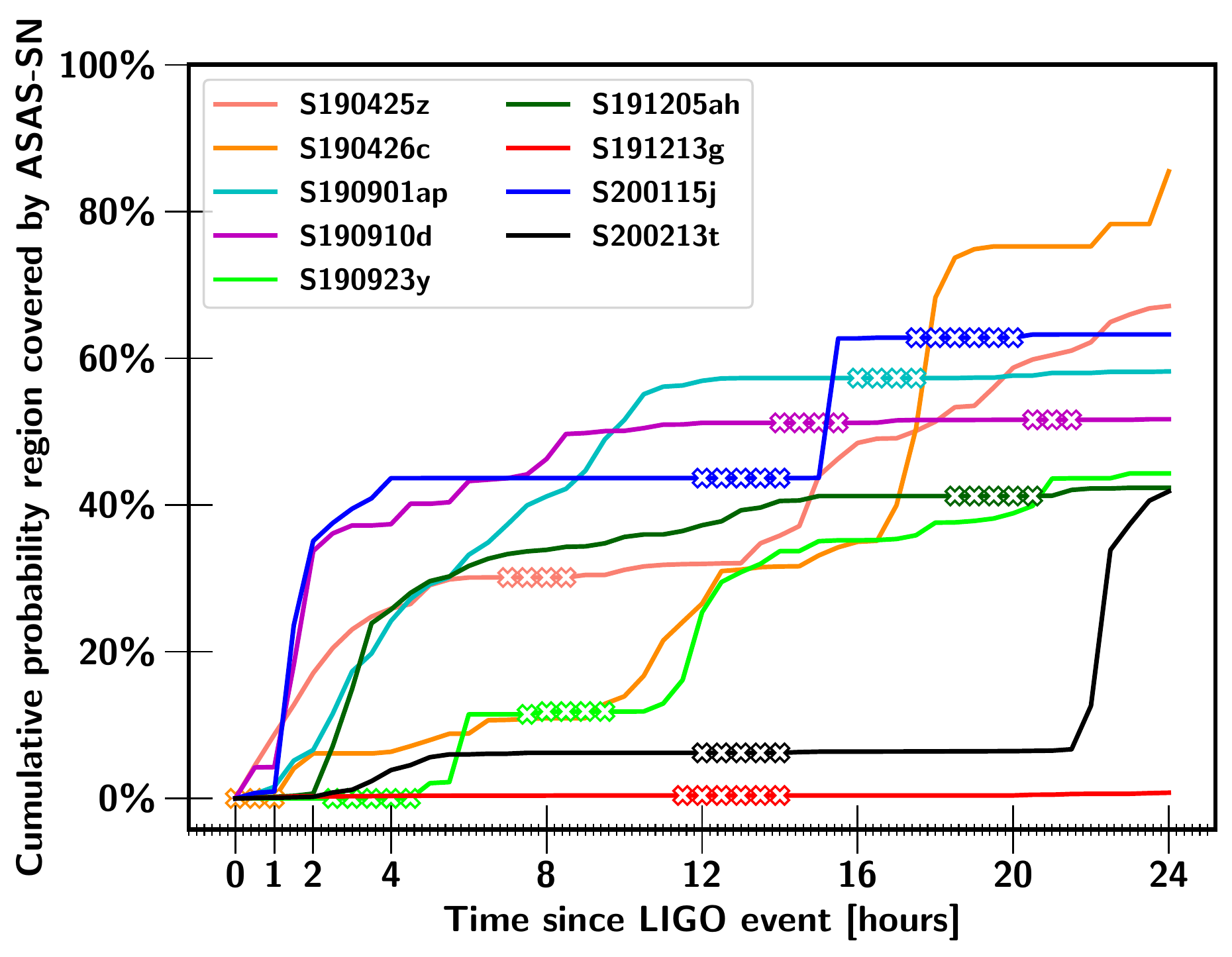}
\caption{Cumulative probability covered by ASAS-SN (all sites) versus the time in hours since the LVC alert for the nine events. The white crosses indicate when a search region is not visible from any ASAS-SN sites due to the restrictions that (1) the Sun is at least 12 degrees below the horizon, (2) the airmass is at most two, (3) the Hour Angle is at most five hours, and (4) that the minimum distance to the Moon is larger than 20$^{\circ}$.
}
\label{fig:proba_time}
\end{figure*}

\subsection{S190425z}
S190425z was the first BNS detection of the LVC O3 run. It was detected on 2019 April 25 at 08:18:05 UTC by the Advanced LIGO-Livingston detector \citep{LVCGCNS190425z} with a false alarm rate of 1 in $\sim$70,000 yr and a $>$ 99\% probability of being a BNS. The first search region was an extended area spanning 10,183 square degrees and the estimated distance was 155 $\pm$ 45 Mpc. Due to the success of the GW170817 optical counterpart hunt and because S190425z was the first event with a high probability of having an optical counterpart, an intense follow-up campaign started a few seconds after the LVC alert, leading to a total of 120 GCN circulars, including two from ASAS-SN \citep{shappee19,shappee19b}.

\begin{figure}
	\includegraphics[width=1.0\columnwidth]{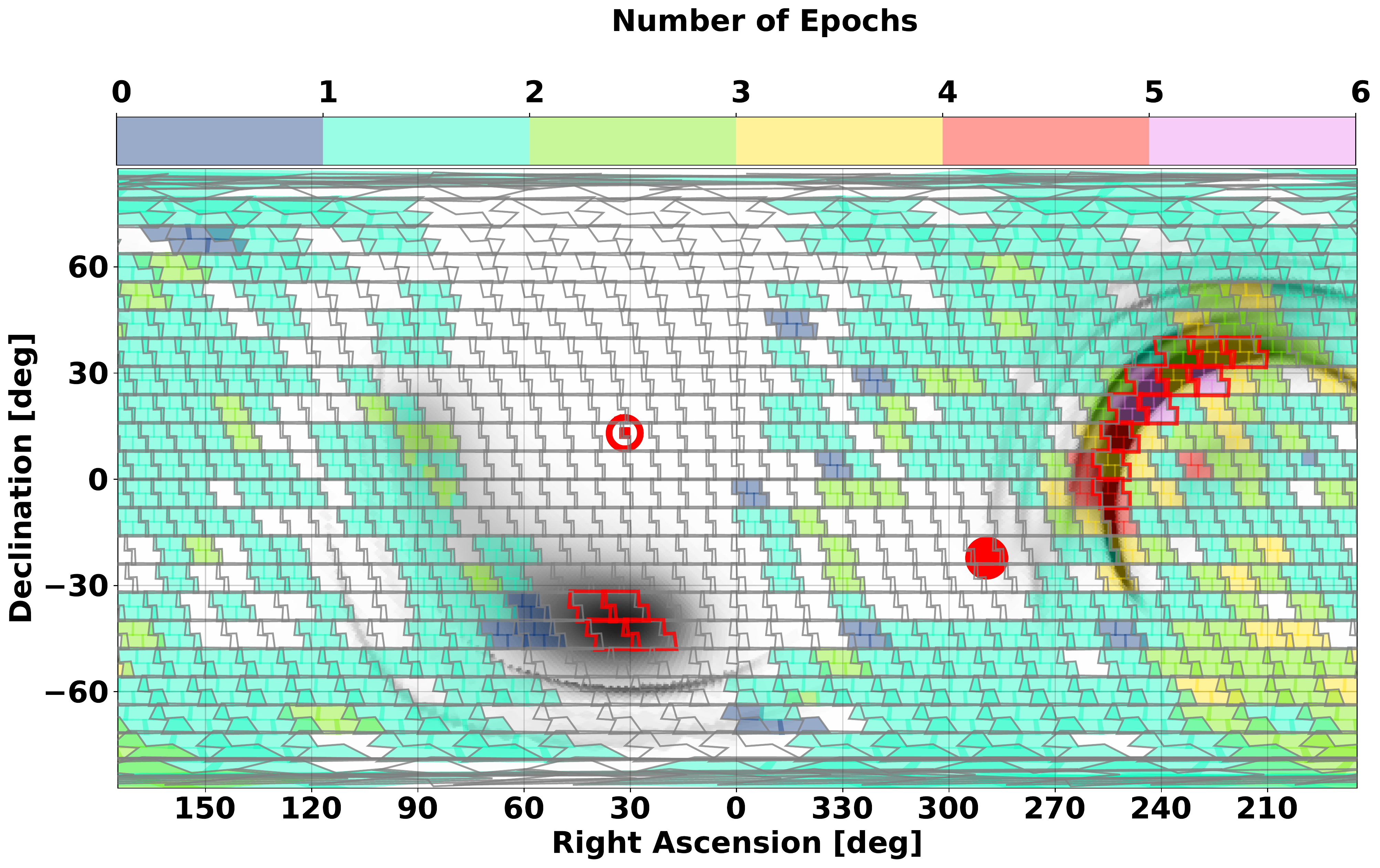}
\caption{Plate-Carree projection of the ASAS-SN data obtained in the first 24 hours for S190425z. The black shaded area corresponds to the localisation area and the red ASAS-SN boundaries are the fields selected to be observed immediately after the event detection. Red filled circle and $\odot$ markers represent the position of the Moon and Sun respectively. The colour of the field shows the number of epochs observed.}
\label{fig:s190425z}
\end{figure}

In Figure \ref{fig:proba_time}, we show the search region covered by ASAS-SN from its five sites. The ASAS-SN response was fast and in less than one hour after the alert, 10\% of the total search region was already mapped through triggered observations to a 5$\sigma$ $g$-band limiting magnitude of $\sim$18.5 mag. After 24 hours, $\sim$70\% of the integrated LVC localisation probability was covered through a combination of normal operations and triggered observations. For comparison, ATLAS covered a sky region totalling of 37.2\% of the event's localisation likelihood in $\sim$6 hours \citep{ATLAS_S190415z}, GOTO covered 29.6\% in 9 hours \citep{goto_S190425z}, Pan-STARRS covered 28\% in $\sim$19 hours \citep{PS_S190425z}, and ZTF covered 21\% in $\sim$28 hours \citep{coughlin20}. Figure \ref{fig:s190425z} shows the ASAS-SN coverage for the 24 hours. As we can see, the normal ASAS-SN observation mode was optimised to increase the cadence for the fields within the search region and we were able to observe those fields 5--6 times over the next day. Finally, in all the search region images taken during the first 24 hours, we flagged 421 stars that had varied, 9 asteroids, and 0 transients.

\subsection{S190426c}

\begin{figure}
	\includegraphics[width=1.0\columnwidth]{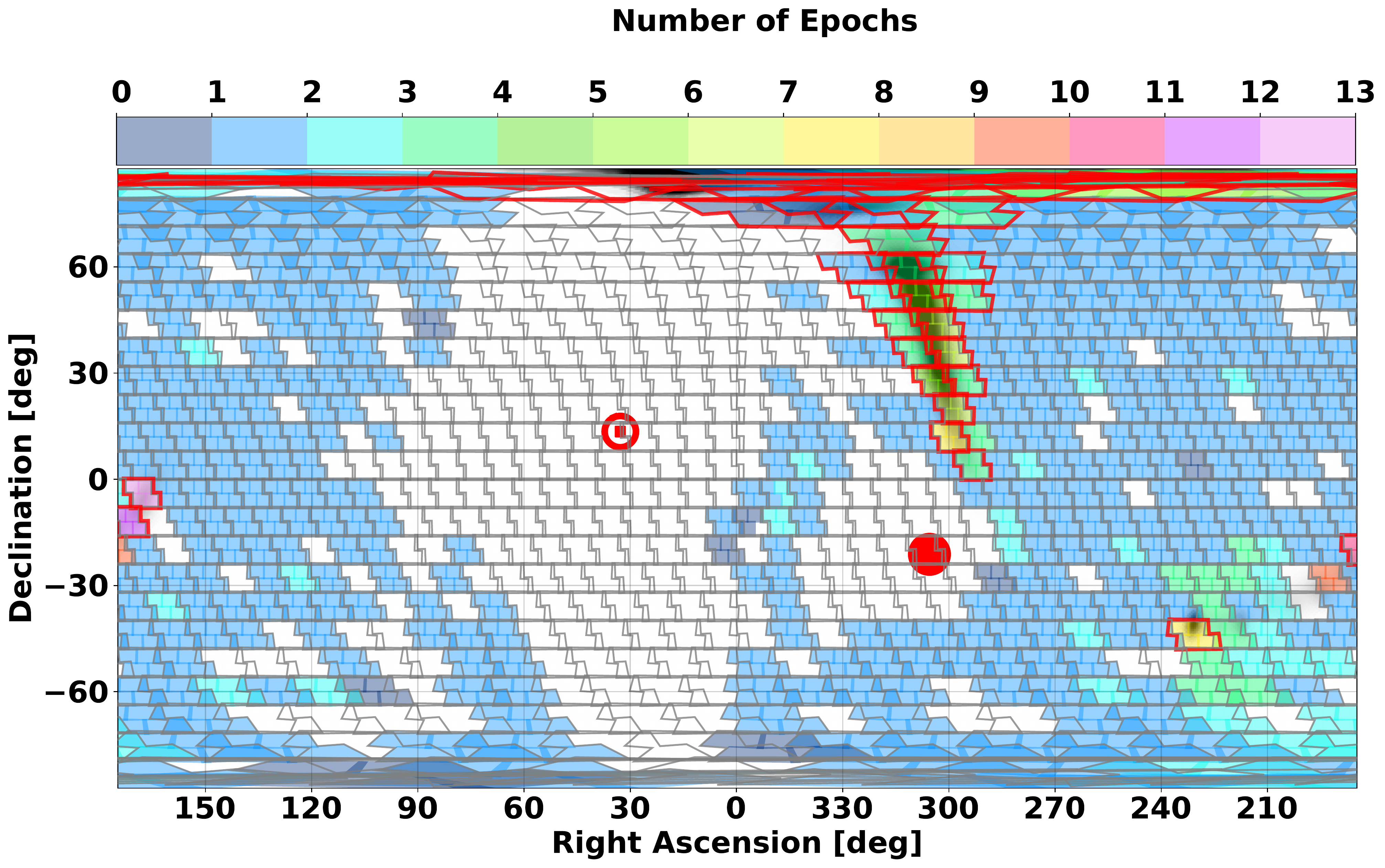}
\caption{Same as Figure \ref{fig:s190425z}, excepted for LVC event S190426c.}
\label{fig:s190426c}
\end{figure}

S190426c was detected by LVC on 2019 April 26 at 15:21:55 UTC with a false alarm rate of 1 in $\sim$1.6 yr \citep{LVCGCNS190426c}. The classification of the GW signal, in order of descending probability, was BNS (49\%), MassGap (24\%), terrestrial (14\%), NSBH (13\%), or BBH (<1\%). MassGap refers to a binary system with at least one compact object whose mass is in the range between NS and BHs, defined as 3--5 ${\rm M}_{\odot}$. The first search region spanned $\sim$1,262 square degrees and the estimated distance was 375 $\pm$ 108 Mpc. Due to the nature of the event, a large number of teams searched for an optical counterpart and $\sim$70 GCN circulars were sent, including one from ASAS-SN \citep{shappee19c}. As seen in Figure \ref{fig:proba_time} and Figure \ref{fig:s190426c}, ASAS-SN mapped $\sim$90\% of the search region in 24 hours. Unlike S190425z, the first fields were observed 1.5 hours after the GW detection due to Sun constraints. Of the nine events presented in this paper, S190426c is the one with the most ASAS-SN observations, with up to 13 epochs over less than 24 hours for the fields within the search region. This demonstrates how we can change the ASAS-SN strategy in real-time and go deeper to discover fainter candidates. For the high cadence fields, we reached a $g$-band limiting magnitude of $\sim$19.8 mag. As with S190425z, ASAS-SN covered one of the largest areas in 24 hours with respect to other major surveys: GOTO covered 54\% of the source location probability in 9 hours but started $\sim 6$ hours after the discovery \citep{goto_S190426c} while ZTF covered 75\% of the sky localisation but in 31 hours and starting 13 hours after the alert \citep{coughlin20}. In all the search region images taken during the first 24 hours, we flagged 1931 variable stars, 79 asteroids, and 1 transient, a cataclysmic variable.

\subsection{S190901ap}

\begin{figure}
	\includegraphics[width=1.0\columnwidth]{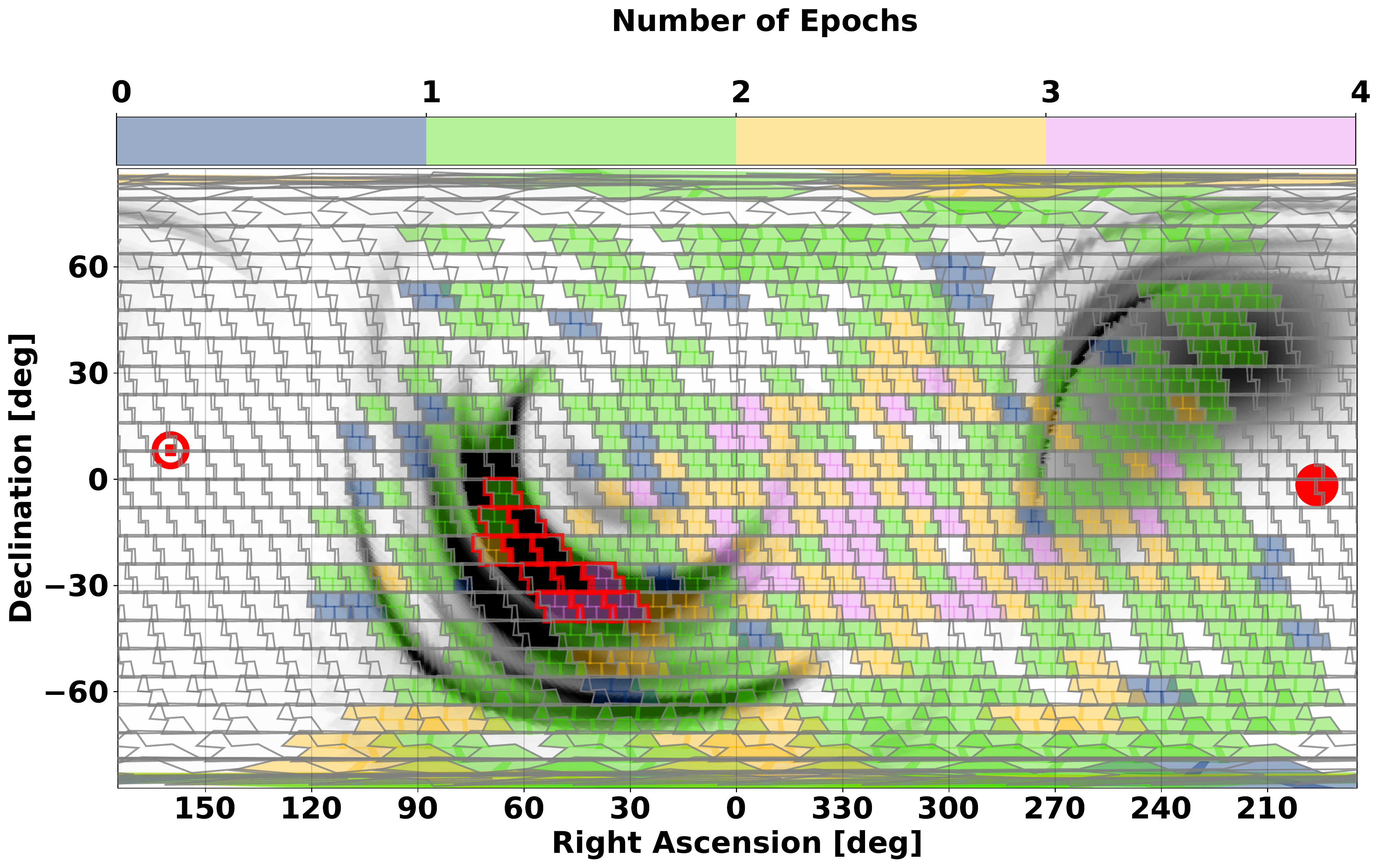}
\caption{Same as Figure \ref{fig:s190425z}, excepted for LVC event S190901ap.}
\label{fig:s190901ap}
\end{figure}

S190901ap was detected on 2019 September 1 at 23:31:01 UTC with a probability of 86\% to be a BNS and 14\% to be terrestrial \citep{LVCGCNS190901ap}. However, its large search region of 13,613 square degrees and distance of 242 $\pm$ 81 Mpc appears to have discouraged groups from searching for an EM counterpart and only $\sim$40 GCN circulars were sent. For this event, the majority of the observed fields within the search region were observed in normal operations, as seen in Figure \ref{fig:s190901ap}, with a median coadded $g$-band depth of $\sim$19.1 mag and some fields reaching $\sim$19.3 mag. Mostly using normal operations, ASAS-SN covered $\sim$60\% of the search region in 24 hours starting $<$ 1 hour after the LVC alert (see Figure \ref{fig:proba_time}). Compared to other teams that reported their observations, ASAS-SN covered one of the largest integrated search probabilities in the first 24 hours. For example, GOTO covered 28\% of the search region in 54 hours starting $\sim 0.1$ hours after the discovery \citep{GOTO_S190901ap} while ZTF covered 73\% of the sky localisation but over $\sim$73 hours and beginning 3.6 hours after the detection \citep{ZTF_s190901ap,coughlin20}. In all the search region images taken during the first 24 hours, we flagged 106 variable stars, 3 asteroids, and 0 transients.

\subsection{S190910d}
\begin{figure}
	\includegraphics[width=1.0\columnwidth]{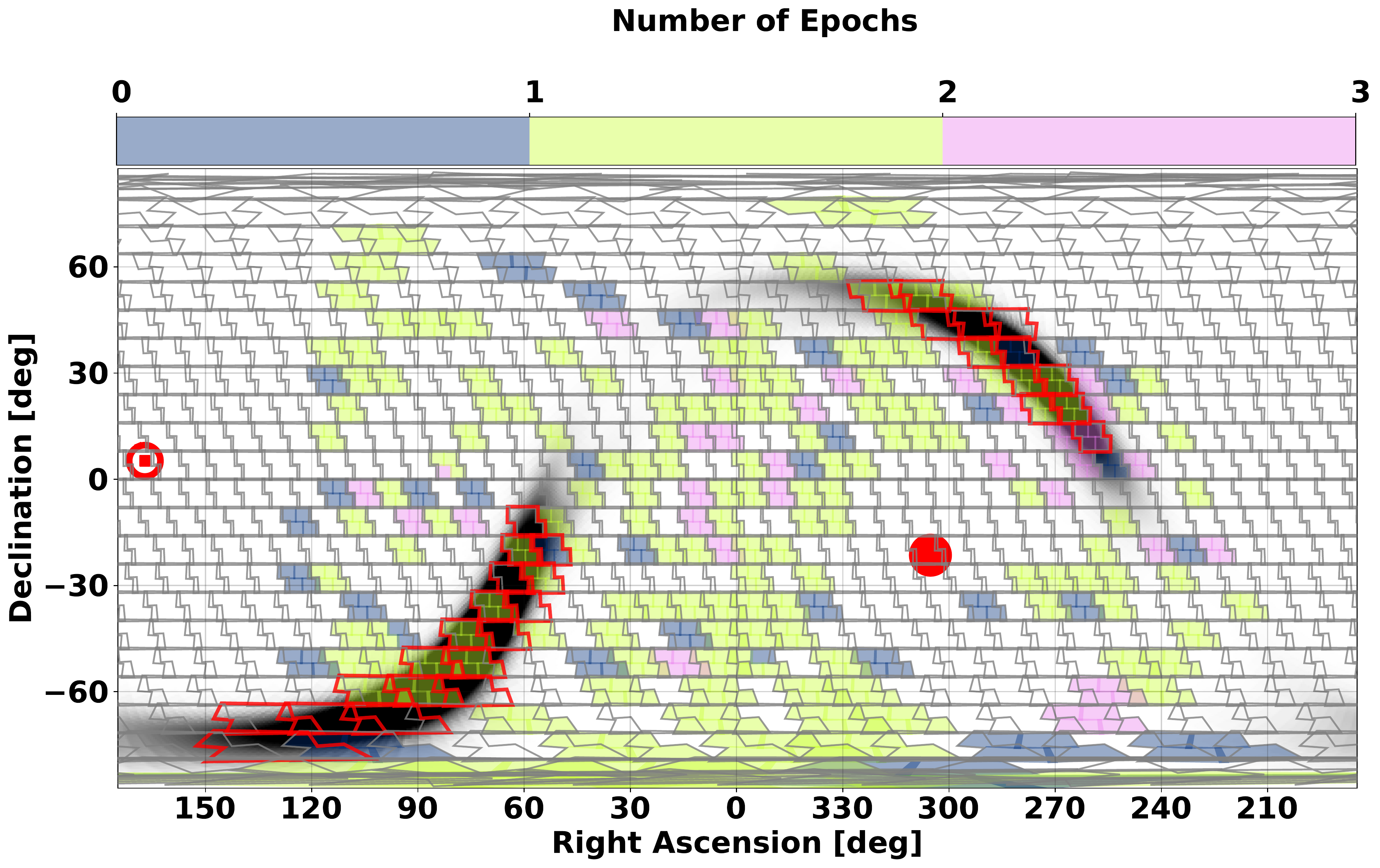}
\caption{Same as Figure \ref{fig:s190425z}, excepted for LVC event S190910d.}
\label{fig:s190910d}
\end{figure}

S190910d was discovered on 2019 September 10 at 01:26:19 UTC \citep{LVCGCNS190910d}. This event was classified as NSBH with a probability of 98\% (terrestrial: 2\%) and a false alert rate of one in eight years. Only a few teams ($\sim$20 GCN circulars) participated in the follow-up of this object situated at 606 $\pm$ 197 Mpc with a search region of 3,829 square degrees. ASAS-SN covered 50\% of the search region in 24 hours (see Figure \ref{fig:proba_time}). For the same object, GRANDMA covered 37\% in 65 hours \citep{grandma_S190910d} while ZTF covered 34\% in 1.5 hours \citep{ZTF_S190910d}. As seen in Figure \ref{fig:s190910d}, ASAS-SN obtained a maximum of three epochs for some fields within the search region and reached a median $g$-band depth of $\sim$17.7 mag with a maximum of $\sim$18.4 mag. In all the search region images taken during the first 24 hours, we flagged 369 variable stars, 0 asteroid, and 0 transients.

\subsection{S190923y}

\begin{figure}
	\includegraphics[width=1.0\columnwidth]{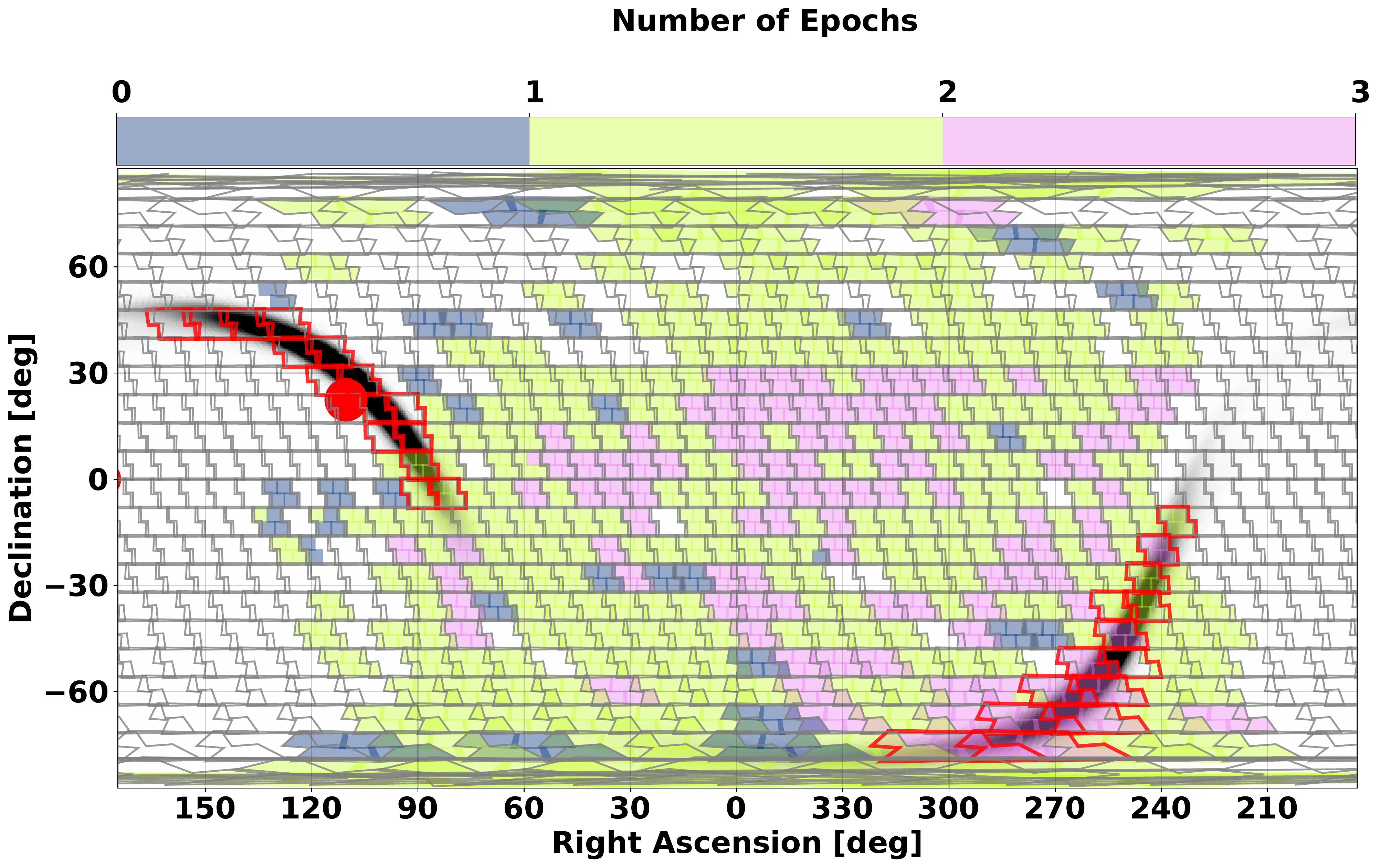}
\caption{Same as Figure \ref{fig:s190425z}, excepted for LVC event S190923y.}
\label{fig:s190923y}
\end{figure}

S190923y was detected on 2019 September 23 at 12:55:59 UTC \citep{LVCGCNS190923y}. The classification for this event was 68\% NSBH, but with a high probability of being terrestrial (32\%). Due to a non-negligible false positive rate of one per seven months, a distance of 438 $\pm$ 151 Mpc, and a search region of 2,107 square degrees, half of which was close to the Moon, only $\sim 15$ GNC circulars were sent. As seen in Figure \ref{fig:s190923y}, ASAS-SN increased the cadence for the fields within the search region and obtained a maximum of three epochs for those fields and a median $g$-band depth of 18.3 mag. As seen in Figure \ref{fig:proba_time}, ASAS-SN covered $\sim$40\% of the search region in 24 hours with the first fields observed starting $\sim$4 hours after the discovery due to observability constraints. As ASAS-SN only targets fields more than 20$^{\circ}$ from the Moon, only half of the search region was observable. In comparison, GRANDMA, covered 26\% in 55 hours starting $\sim$4 hours after the alert \citep{grandma_S190923y}. In all the search region images taken during the first 24 hours, we flagged 2084 variable stars, 47 asteroids, and 0 transients.

\subsection{S191205ah}
\begin{figure}
	\includegraphics[width=1.0\columnwidth]{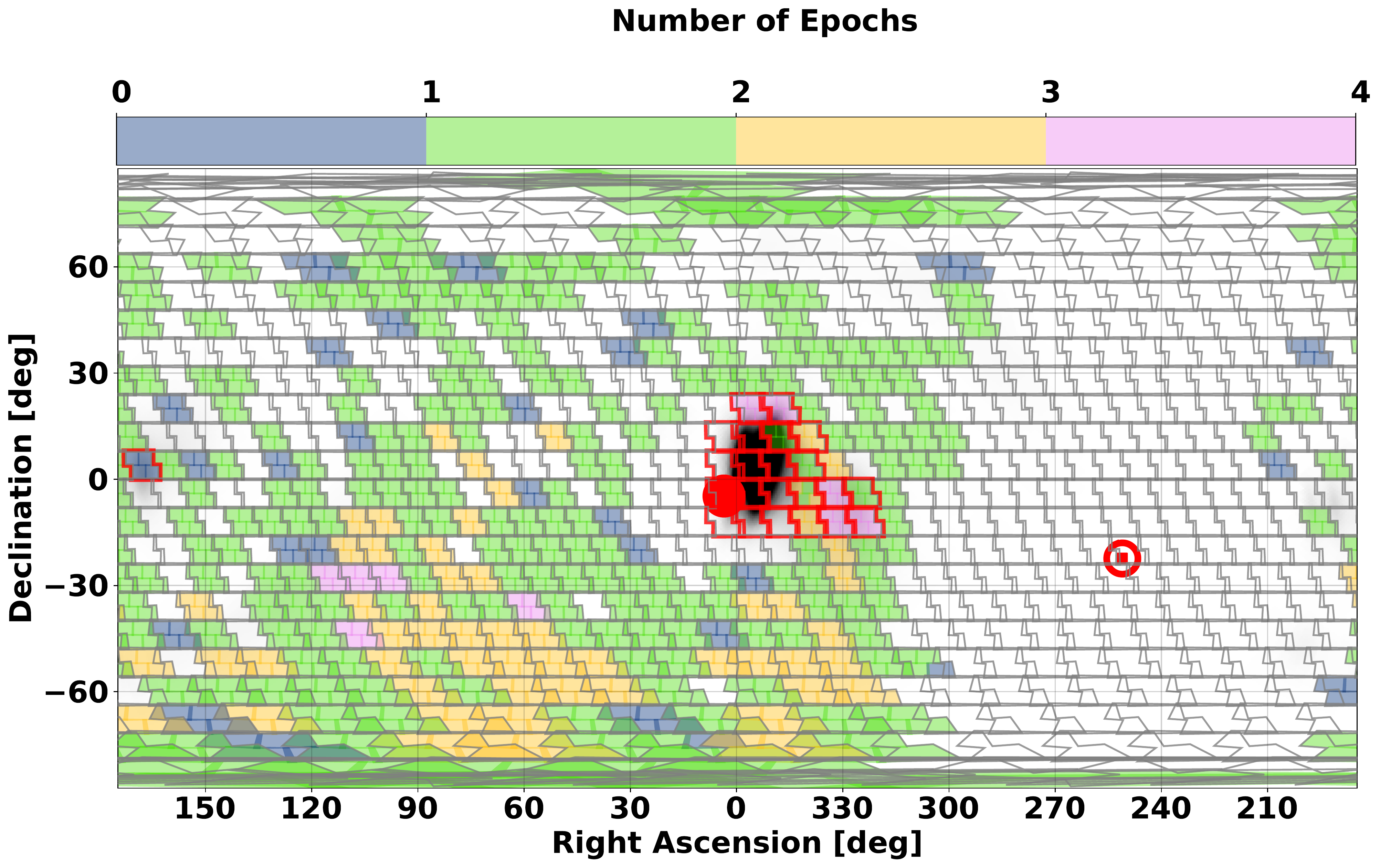}
\caption{Same as Figure \ref{fig:s190425z}, excepted for LVC event S191205ah.}
\label{fig:s191205ah}
\end{figure}

S191205ah was detected on 2019 December 5 at 21:52:08 UTC \citep{LVCGCNS191205ah} with a false alarm rate of $\sim$1 per 2.5 years. The classification was 93\% NSBH and 7\% terrestrial at a distance of 385 $\pm$ 164 Mpc. The GW candidate was poorly localised with a search region spanning 6,378 square degrees. Only a few teams participated in the EM counterpart hunt and only $\sim$30 GCN circulars were sent. Through its routine and optimised survey operations, ASAS-SN covered $\sim$40\% of the search region in 24 hours (see Figure \ref{fig:proba_time}) with the first observation 2 hours after the discovery. ZTF covered 6\% in 167 hours starting $\sim$11 hours post-discovery \citep{ztf_S191205ah,coughlin20b} and GRANDMA mapped 4.8\% in $\sim$150 hours \citep{antier20}. Figure \ref{fig:s191205ah} shows that our strategy obtained multiple epochs for the high probability regions that were sufficiently distant from the Moon. The median $g$-band depth was 16.97 mag with a maximum of 18.30 mag. In all the search region images taken during the first 24 hours, we flagged 101 variable stars, 37 asteroids, and 0 transients.

\subsection{S191213g}

\begin{figure}
	\includegraphics[width=1.0\columnwidth]{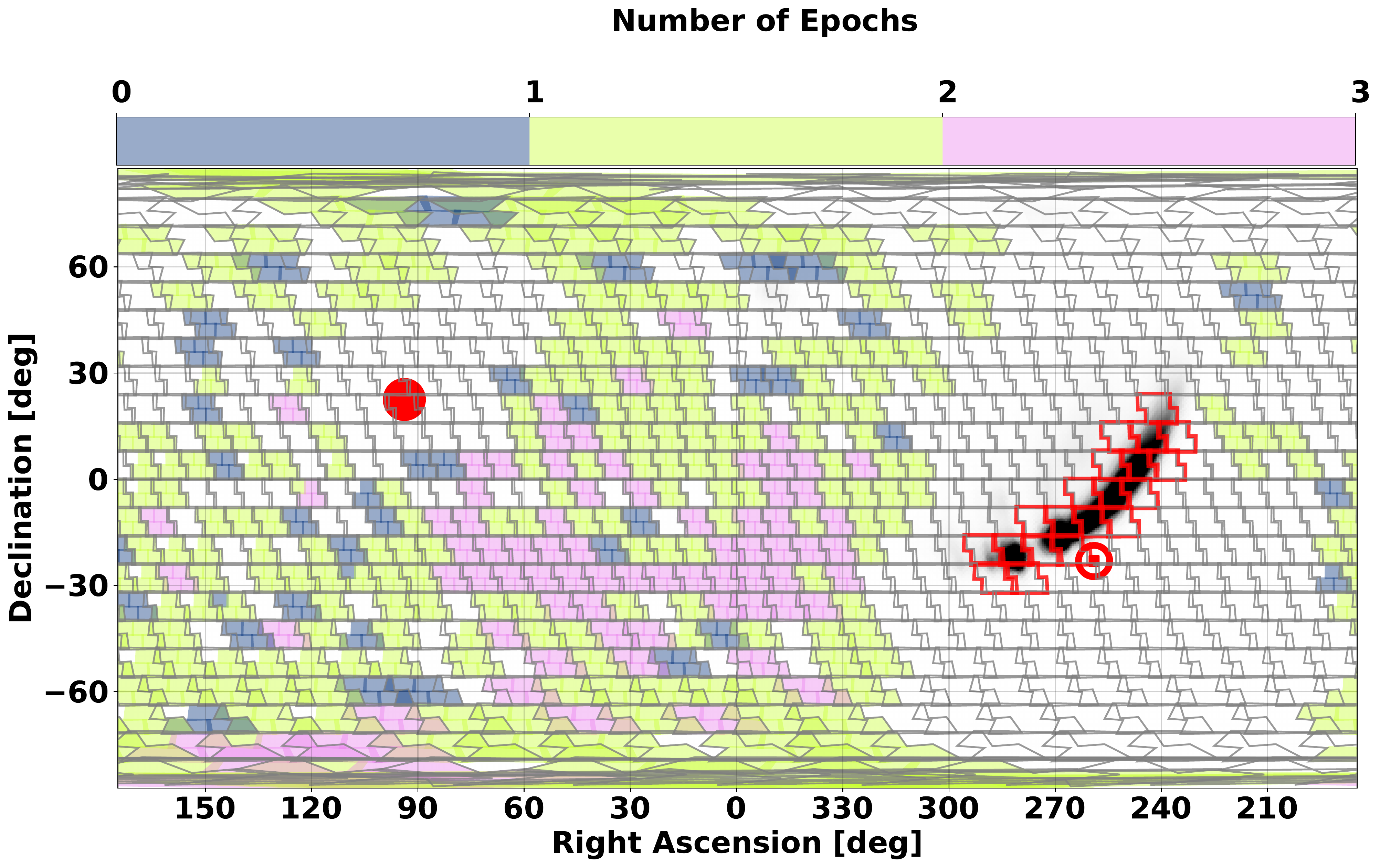}
\caption{Same as Figure \ref{fig:s190425z}, excepted for LVC event S191213g. This candidate was not observable due to our Sun and our Hour Angle limits.}
\label{fig:s191213g}
\end{figure}

S191213g was detected on 2019 December 12 at 04:34:08 UTC \citep{LVCGCNS191213g}. The classification probabilities were 77\% BNS and 23\% terrestrial. With a false alarm rate of 1.12 per year, a distance of 195 $\pm$ 59 Mpc, and a search region of 1393 square degrees, S191213g was fairly well observed ($\sim$50 GCN circulars). However, as seen in Figure \ref{fig:s191213g}, the search region was within the Sun and Hour Angle limit of five hours so ASAS-SN did not observe it, leading to its 0\% coverage in Figure \ref{fig:proba_time}. ZTF was able to cover 28\% of the probable region in 27 hours by observing at high airmass in twilight \citep{ztf_s191213g} while GRANDMA observed only 1\% over 73 hours \citep{grandma_S191213g}.

\subsection{S200115j}

\begin{figure}
	\includegraphics[width=1.0\columnwidth]{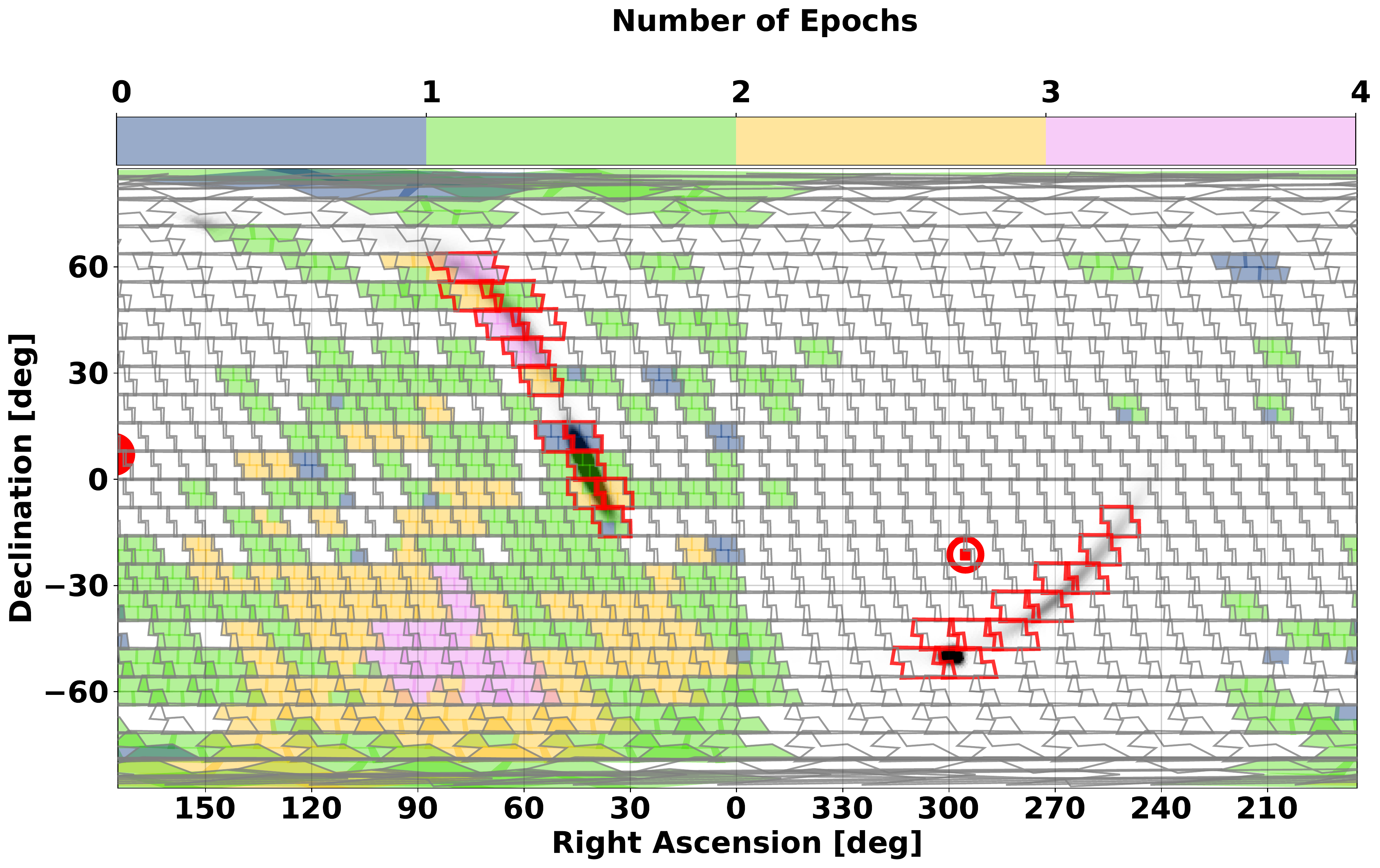}
\caption{Same as Figure \ref{fig:s190425z}, excepted for LVC event S200115j.}
\label{fig:s200115j}
\end{figure}

S200115J was identified on January 15, 2020, at 04:23:08 UTC with a false alarm rate of 1 per 1,573 years \citep{LVCGCNS200115j}. It was classified as a MassGap event with a probability of 99\%, a distance of 331 $\pm$ 97 Mpc, and a search region spanning 908 square degrees, one of the smallest regions in LVC O3. As seen in Figure \ref{fig:s200115j}, only two-thirds of the search region was observed by ASAS-SN due to Sun constraints. Fields within the search region were targeted at a slightly higher frequency (2--4 epochs) with a median $g$-band depth of 17.8 mag and a maximum of 18.5 mag. For this event, ASAS-SN was among a few groups to obtain data and only $\sim$30 GCN circulars were sent. As shown in Figure \ref{fig:proba_time}, ASAS-SN was able to cover $\sim$60\% of the search region in 24 hours and the first fields within the search region were observed 1 hour after the LVC alert. Among other teams, SAGURO observed 2.1\% of the LVC total probability in 24 hours starting 72 hours after the event \citep{paterson21}, ZTF scanned 22\% in one hour to a depth of 21.2 mag starting 15 minutes after the alert \citep{ztf_s200115j,coughlin20b}, and GRANDMA observed 7\% in 24 hours starting 11 hours after the event with a depth of 17 mag in clear band. In all the search region images taken during the first 24 hours, we flagged 22 variable stars, 16 asteroids, and 0 transients.

\subsection{S200213t}

\begin{figure}
	\includegraphics[width=1.0\columnwidth]{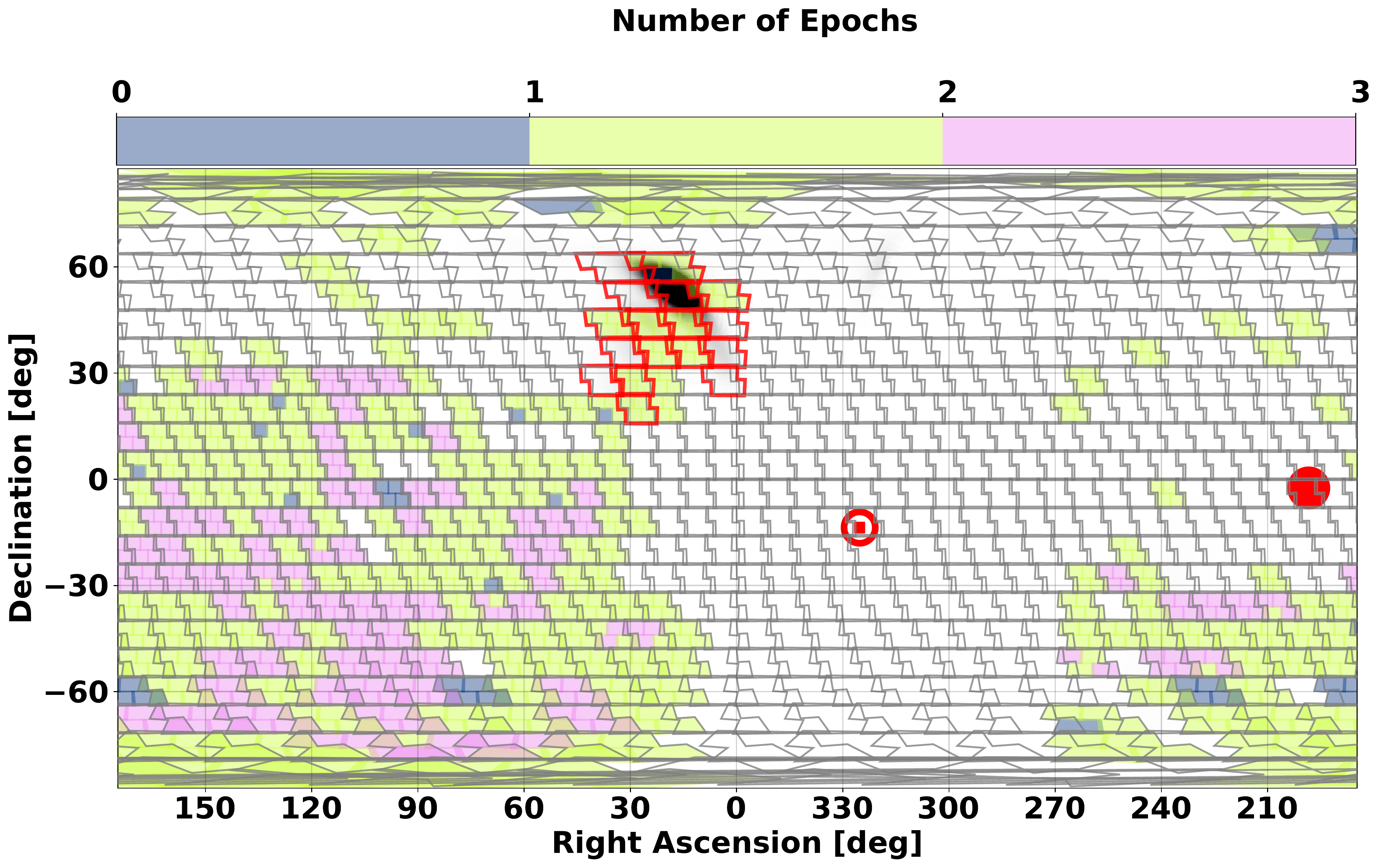}
\caption{Same as Figure \ref{fig:s190425z}, excepted for LVC event S200213t.}
\label{fig:s200213t}
\end{figure}

The last GW candidate with a high NS probability was discovered on February 13, 2020, at 04:10:40 UTC \citep{LVCGCNS200213t}. S200213t had a false alarm rate of $\sim$1 per 1.8 years, a search region of 2,587 square degrees, and a distance of 224 $\pm$ 90 Mpc. For this event, the probability of a terrestrial origin was non-negligible (37\%), but due to the 63\% probability of being an BNS, several searches followed up this signal ($\sim$50 GCN circulars). The localisation region was confined to relatively high northern latitudes, and the two northern ASAS-SN sites were not immediately able to obtain data. Haleakala Observatory was closed due to damage from an ice storm and McDonald Observatory had bad weather during the $\sim$2 hour ToO period. As we can see in Figure \ref{fig:s200213t}, all the fields overlapping the search region were ultimately observed. A maximum of 2--3 epochs were obtained with a median $g$-band depth of 18.5 mag. ASAS-SN was able to cover $\sim$40\% of the search region, mostly between 22 and 24 hours after the event. In comparison, ZTF covered 72\% of the area in less than 1 day to a $g$-band magnitude limit of $\sim$20.7 mag \citep{ztf_s200213t}, GOTO observed 54\% in $\sim$26 hours starting a few seconds after the event, and GRANDMA mapped $\sim$30\% in $\sim$44 hours starting 30 min after the event. In all the search region images taken during the first 24 hours, we flagged 208 variable stars, 11 asteroids, and 1 transient, a luminous red nova or a luminous blue variable \citep{19zhd_tns}.

\section{Discussion}\label{txt:discussion}

In Section \ref{txt:events}, we detailed all the GW candidates with a high probability of having an optical counterpart. Like the rest of the community, we did not detect a kilonova during LVC O3. To understand why, we will first study the case of SSS17a, the only kilonova detected to date. Then we will use SSS17a-like objects as a tool to improve our observational strategy for detecting future kilonovae.

\subsection{S170817a}

Unfortunately, when S170817a was discovered, all the LVC alerts were private and ASAS-SN was not part of the collaboration. For this reason, ASAS-SN covered 0\% of the search region and the only data obtained were gathered in the normal observation mode. Due to the small 28 square degree size of the search region, it was contained in just five pointings and our likelihood of having randomly observed this event were small for two reasons. First, SSS17a was only observable early in the night and fields far over in either direction are less likely to be observed due to the limited amount of time available to observe them. Second, at that time, only the Brutus and Cassius sites were operational. Paczynski, Leavitt, and Payne-Gaposchkin  were just coming online, obtaining their first night of data a few weeks after the event on September 2nd, September 20th, and October 29th, 2017, respectively.

\begin{figure}
	\includegraphics[width=1.0\columnwidth]{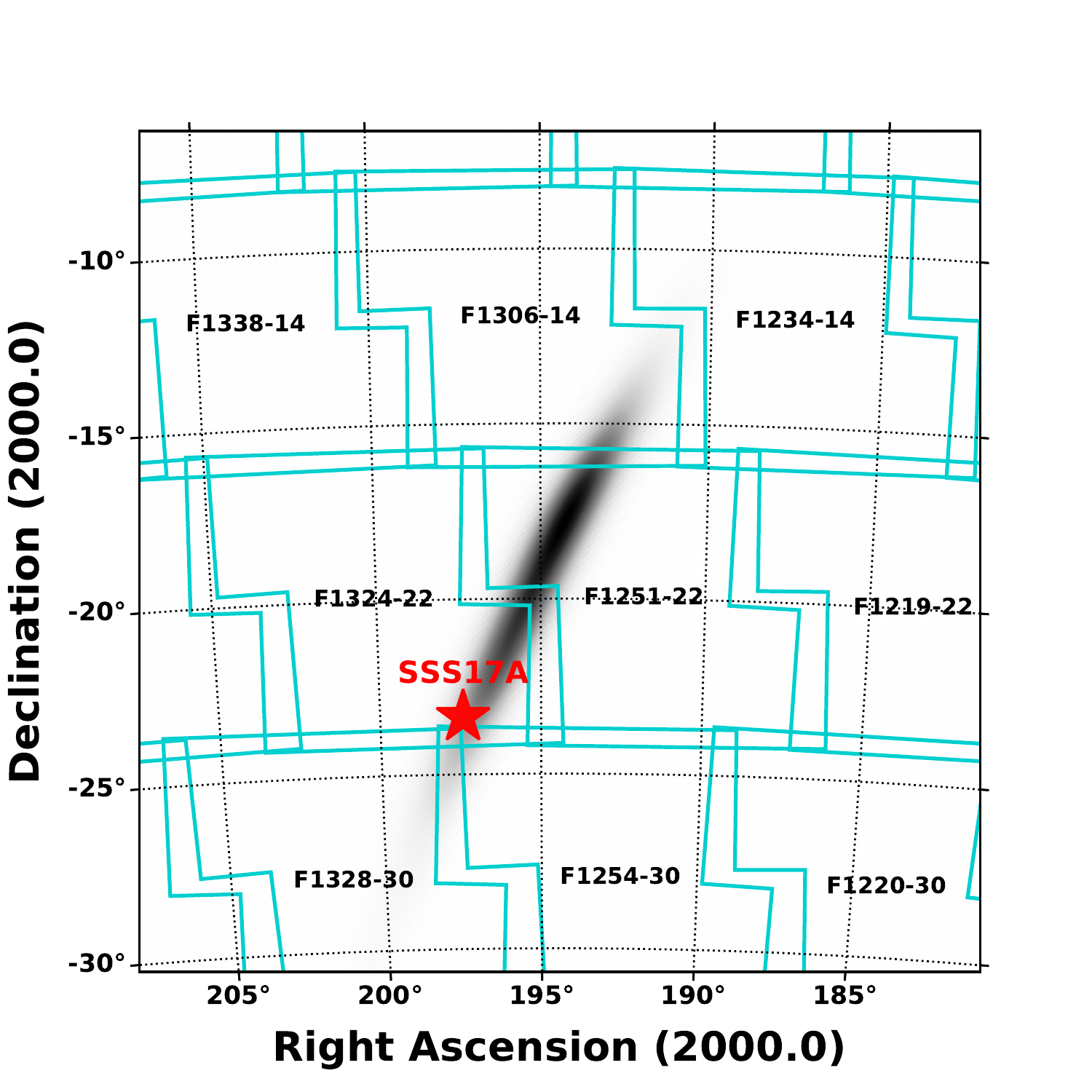}
\caption{Hammer Aitoff projection of the ASAS-SN fields (cyan) with the black shaded area corresponding to the search region of SSS17a. The red star represents the position of SSS17a. The names of the ASAS-SN fields are also written in black.}
\label{fig:sss_17a_field}
\end{figure}

However, given that LVC alerts are now public, all five sites are available, and we have observationally tested our initial strategy, we can use SSS17a-like objects to investigate how ASAS-SN should respond to future kilonova events. Figure \ref{fig:sss_17a_field} shows a Hammer Aitoff projection of the ASAS-SN fields superposed on the search region for SSS17a. Five pointings cover 98\% of the search region and just two pointings (F1251$-$22 and F1324$-$22) cover 70\% of the total probability. Given our strategy as described in Section \ref{sec:strategy}, ASAS-SN would trigger observations of the five fields by order of probability (i.e., F1251$-$22, F1324$-$22, F1306$-$14, F1328$-$30, and F1254$-$30). This means that we would have discovered SSS17a a few seconds after the LVC alert in our second pointing, $\sim$10 min after starting the hunt. For comparison, \citet{coulter17} -- who were the first team to discover SSS17a $\sim$11 hours after the alert -- did a galaxy-targeted search with the Swope telescope (Las Campanas Observatory, Chile) and found SSS17a in their ninth image (the twelfth-highest probability galaxy). This means that ASAS-SN would have discovered SSS17a almost at the same time as \citet{coulter17} using only the Cassius site in Chile. However, with the ASAS-SN unit in South Africa (Payne-Gaposchkin), deployed two months after GW170817, ASAS-SN would have discovered SSS17a 4.5 hours after the LVC alert instead of 10--11h from Chile. Such early data are crucial to differentiate shock cooling and radioactively powered models and to explain the blue component of the kilonova \citep{piro18}. It is also important to note that unlike the majority of other surveys, ASAS-SN already has deep templates over the entire sky and a $\sim$10 years of variability history.

\begin{figure}
	\includegraphics[width=1.0\columnwidth]{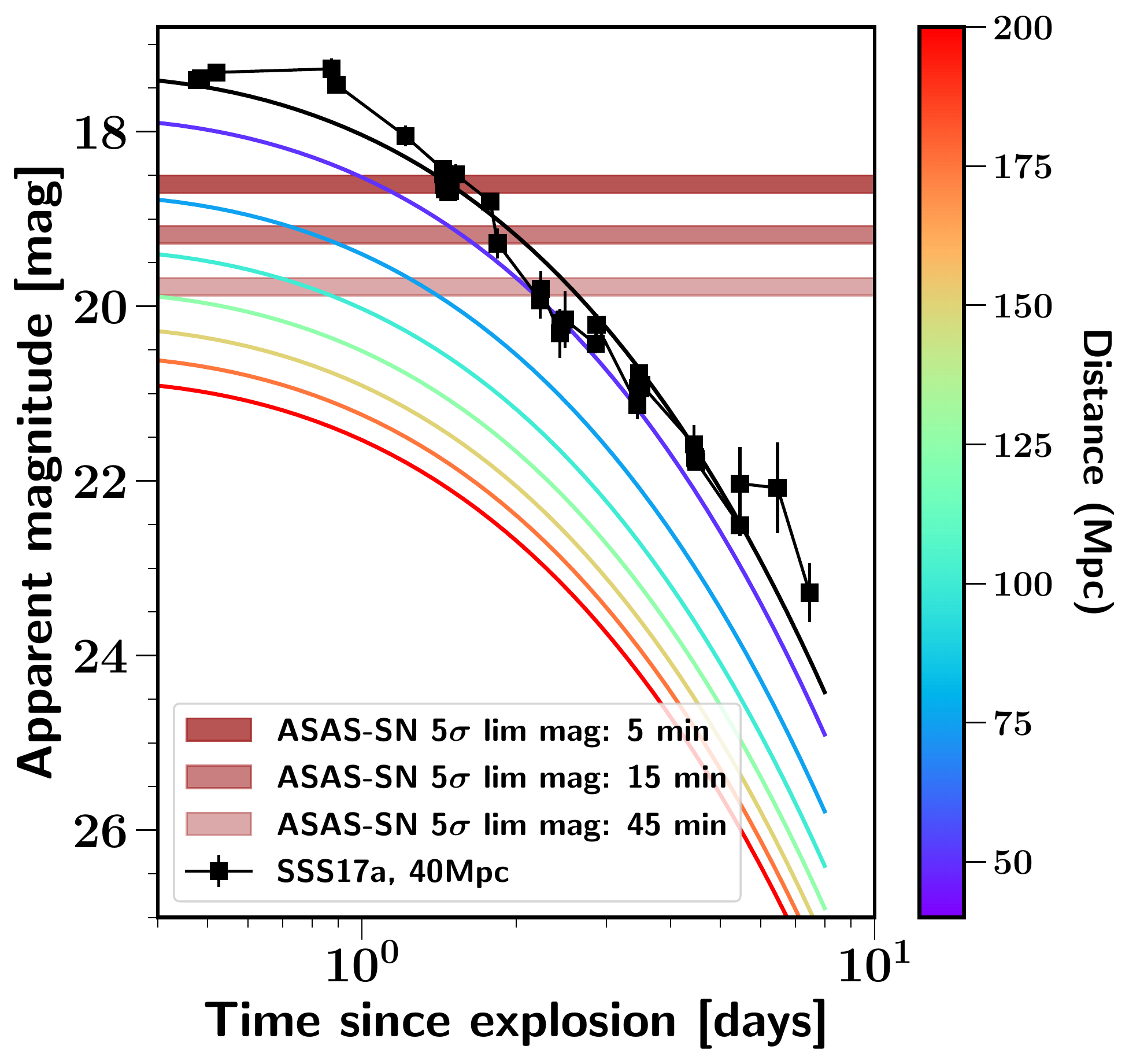}
\caption{SSS17a $g$-band apparent magnitude evolution for different distances. The original light curve is displayed in black (40 Mpc) while the lines corresponds to the cocoon model \citep{piro18} for different distances in Mpc defined by the colour bar. The three brown horizontal bands are the 5$\sigma$ $g$-band limiting magnitude from ASAS-SN for exposures time per field of 5, 15, and 45 minutes respectively (see Section \ref{txt:LVC04} for more details). The $g$-band data were taken from the compilation in \citet{villar17} and were originally published in \citet{andreoni17}, \citet{arcavi17}, \citet{coulter17}, \citet{cowperthwaite17}, \citet{diaz17}, \citet{drout17}, \citet{pian17}, \citet{shappee17}, \citet{smartt17}, \citet{troja17}, and  \citet{utsumi17}.}
\label{fig:sss17a_mag}
\end{figure}

SSS17a was easily discovered mostly thanks to its small distance (40 Mpc). In Figure \ref{fig:sss17a_mag}, we show the SSS17a $g$-band light curve \citep{andreoni17,arcavi17,coulter17,cowperthwaite17,diaz17,drout17,pian17,shappee17,smartt17,troja17,utsumi17} not corrected for Milky Way extinction and the cocoon model from \citet{piro18} for distances from 40 Mpc to 200 Mpc. With ASAS-SN and an average 5$\sigma$ $g$-band limiting magnitude of 18.5 mag for a single epoch per field ($\sim$5 minutes), SSS17a was bright enough for ASAS-SN to detect for $\sim$1.5 days even with our normal survey observations. The maximum distance for a single normal survey epoch to detect an SSS17a-like source is $\sim$60--75 Mpc. However, once triggered we also increase the cadence for the fields within the source region, leading to a fainter 5$\sigma$ limiting magnitude. For example, for small probability areas, it would be easy to reach 45 minutes of exposure time per field corresponding to a limiting magnitude of $\sim$19.6 mag and the ability to detect SSS17a at a distance of $\sim$120 Mpc (see also Section \ref{txt:LVC04}).

SSS17a suffered from a higher than average Milky Way $g$ band extinction of $\sim$0.4 mag \citep{schlafly11}. As more than half of the sky has a lower Milky Way extinction, most future kilonovae will be less extinguished and detectable at moderately larger distances.

\subsection{Strategy for LVC O4}\label{txt:LVC04}

The next LVC observing run is expected to start in June 2022 with an additional detector located in Japan (Kagra). LIGO-Virgo-Kagra (LVK) detections have a predicted median search area of 33$^{+4.9}_{-5.3}$ square degrees \citep{abbott20b}, an area similar to SSS17a. 

\begin{figure}
	\includegraphics[width=1.0\columnwidth]{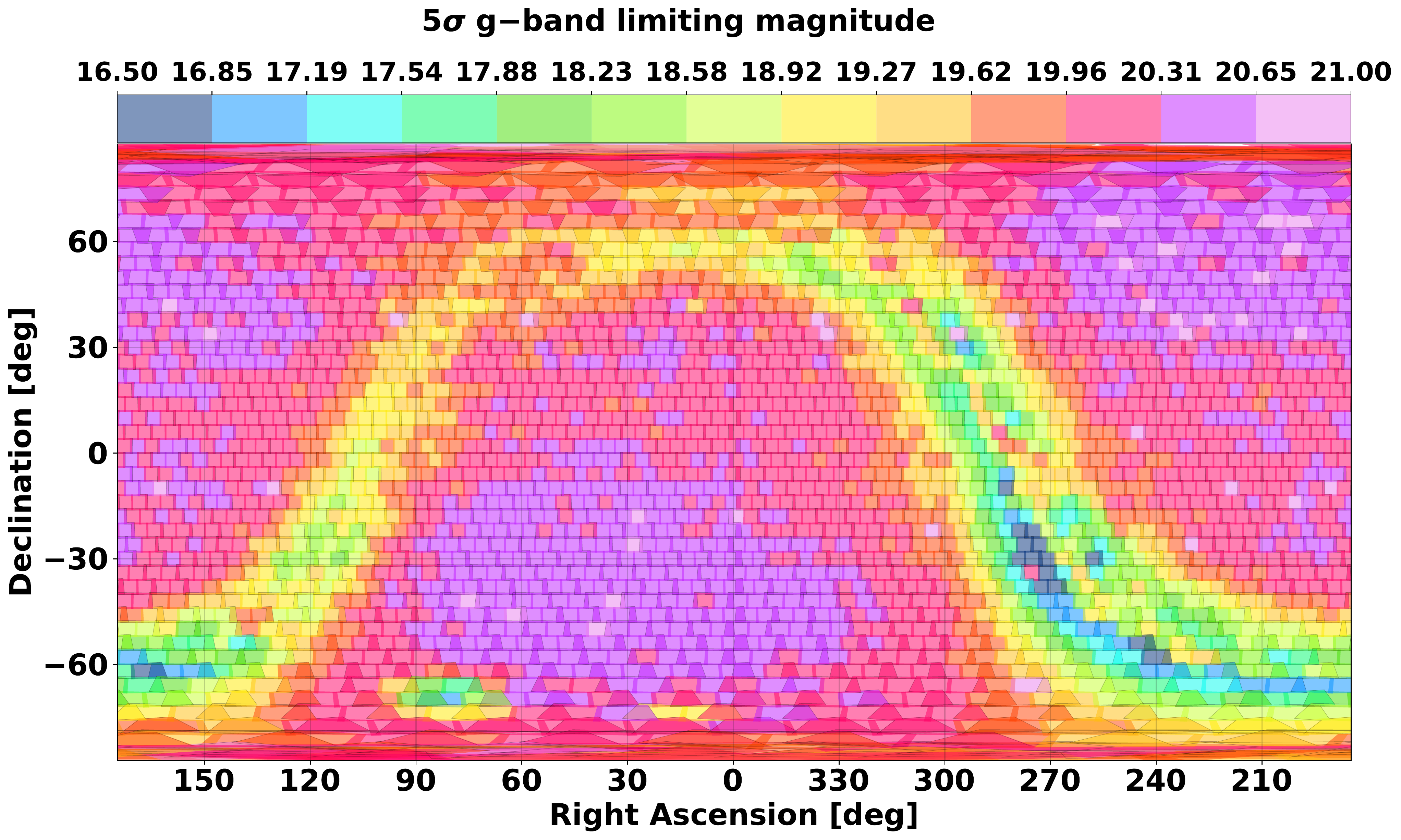}
\caption{Plate-Carree projection of the depth of ASAS-SN $g$-band reference images over the sky. The colour of the different ASAS-SN field shows the 5$\sigma$ $g$-band limiting magnitude of the reference images. The median depth over all the sky is 20.3 mag.}
\label{fig:mag_lim}
\end{figure}

\begin{figure}
	\includegraphics[width=1.0\columnwidth]{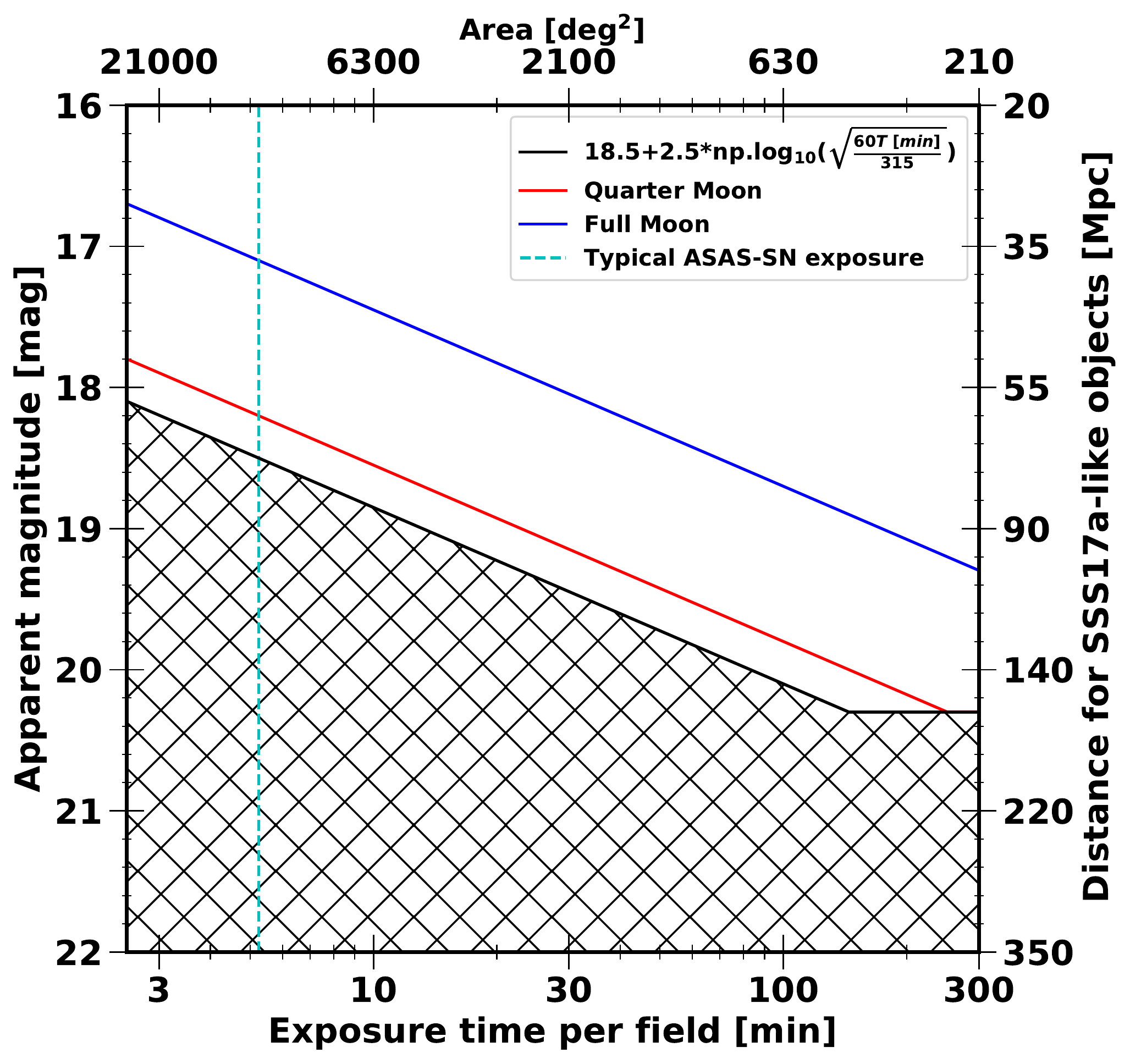}
\caption{Apparent magnitude limits versus total exposure time in minutes.
The black, red, and blue lines represent are 5$\sigma$ $g$-band limiting
magnitude for different moon phases: New, Quarter, and Full Moon. The right axis is the distance at which SSS17a would be detected given an apparent magnitude. The top axis is the search area that ASAS-SN can cover for a given exposure time per field using 3 of the 5 ASAS-SN sites and assuming fields are visible five hours per night. If a candidate lies in the hatched region, ASAS-SN should not trigger based on SSS17a. The vertical cyan dashed line is the ASAS-SN exposure time in normal operations.}
\label{fig:strategy_04}
\end{figure}

Because ASAS-SN already has deep reference images covering the entire sky, we can obtain longer exposure times to search for counterparts at fainter magnitudes without significant noise contribution from the reference image. Figure \ref{fig:mag_lim} shows the depth of ASAS-SN $g$-band reference images. A typical reference image is comprised of images totalling several hours of exposure time, with a median 5$\sigma$ $g$-band limiting magnitude of 20.3 mag. Figure \ref{fig:strategy_04} illustrates which events ASAS-SN should follow given its distance and search area, assuming a peak absolute magnitude like SSS17a. We assume the event is visible for five hours from three of our five sites and that SSS17a stayed near its peak for 300 minutes (see Figure \ref{fig:sss17a_mag}). 

For simplicity, we ignore edge effects and assume that the total field of view of ASAS-SN is entirely contained within the search region, so that each pointing covers 70 square degrees of the search region. For the largest search areas, we can only dedicate $\sim$5 minutes to each field to cover $\sim$12,600 square degrees in one night at a depth of 18.5 mag. This would detect a SSS17a out to $\sim$70 Mpc. If we observe only one field, we can reach a total exposure time of 900 minutes (3 sites $\times$ 5 hours) in one night. For these very long exposures, the sensitivity will be limited by the depth of the reference images and we can reach a 5$\sigma$ $g$-band limiting magnitude of 20.3 mag to detect SSS17a-lke object out to $\sim$160 Mpc. Of course, by the start of the next LVC run, the reference images will be deeper than shown in Figure \ref{fig:mag_lim}.

These restrictions mean we should only follow GW candidates with a distance $\leq$1.5 $\times$ 10$^{2.87-0.5 \rm{log}_{10}(\sqrt{A})}$ Mpc for a search region of $A$ square degrees. This limit is conservatively 50\% deeper than the one represented by the black line in Figures \ref{fig:strategy_04}, as SSS17a is only one example of a kilonova. With this new strategy, the majority of the events from LVC O3 would not have been triggered as they were too distant or their skymap areas were too large. Only three events, S190415z, S191213g, and S200213t, with minimum distances of 110, 136, and 134 Mpc, respectively, would have been observed by ASAS-SN under this strategy.

\section{Conclusions}\label{txt:conclusions}

We described the ASAS-SN searches for EM counterparts of GW candidates detected by the Advanced LIGO/Virgo during the third science run (April 2019--February 2020). We observed a total of nine events, all with a $>$60\% probability of having at least one NS component. Unfortunately, like the rest of the community, ASAS-SN did not detect an optical counterpart. However, thanks to its large field of view and its five units located at four sites in both hemispheres, ASAS-SN covered an average of 50\% of the LVC search region in less than 24 hours, starting a few minutes after the alert for some events. For seven events, ASAS-SN is the survey that covered the most area in 24 hours with a $g$-band limiting magnitude of 18.5 mag. We were also able to trigger and observe the fields in the search region at a higher cadence, obtaining up to a dozen epochs within 24 hours even for some of the larger search region. For LVC O4, we will only trigger on closer or well-localised sources: events where the distance D is smaller than $\sim$1.5 $\times$ 10$^{2.87-0.5 \rm{log}_{10}(\sqrt{A})}$ Mpc for a given search region spanning $A$ square degrees. With our new observational strategy, we will triggeron fewer objects but with a higher cadence and deeper observations.

With the additional Kagra detector and the LVC detector upgrades, the predicted rate of BNS merger detections will increase to 10$^{+52}_{-10}$ per year from 1$^{+12}_{-1}$ for O3 \citep{abbott20b} and the median search area will decrease to 33$^{+4.9}_{-5.3}$ square degrees \citep{abbott20b}, roughly the same search area as for SSS17a. Even for an elongated shape, ASAS-SN should be able to map the full search region in only 2--5 pointings to a $g$-band depth of $\sim$18.5 mag, taking only five minutes per field. With more epochs/exposure time, we will be able to reach a limiting magnitude of $\sim$20.3 mag and detect candidates at distances up to $\sim$160 Mpc assuming an SSS17a-like light curve. With LVK and our new observational strategy, ASAS-SN will play a key role in searching bright optical counterparts during the LVC O4 run and one of the few able to observe both hemispheres.

\section*{Acknowledgements}
Support for T.d.J has been provided by NSF grants AST-1908952 and AST-1911074. B.J.S. is supported by NSF grants AST-1907570, AST-1908952, AST-1920392, and AST-1911074. CSK and KZS are supported by NSF grants AST-1814440 and AST-1908570. Support for TW-SH was provided by NASA through the NASA Hubble Fellowship grant HST-HF2-51458.001-A awarded by the Space Telescope Science Institute, which is operated by the Association of Universities for Research in Astronomy, Inc., for NASA, under contract NAS5-265. J .F.B. is supported by National Science Foundation grant No.\ PHY-2012955.

ASAS-SN is funded in part by the Gordon and Betty Moore Foundation through grants GBMF5490 and GBMF10501
  to the Ohio State University, NSF grant AST-1908570, the Mt. Cuba Astronomical Foundation, the Center for Cosmology and AstroParticle Physics (CCAPP) at OSU, the Chinese Academy of Sciences South America Center for Astronomy (CAS-SACA), and the Villum Fonden (Denmark). Development of ASAS-SN has been supported by NSF grant AST-0908816, the Center for Cosmology and AstroParticle Physics at the Ohio State University, the Mt. Cuba Astronomical Foundation, and by George Skestos. Some of the results in this paper have been derived using the healpy and HEALPix packages.

This work is based on observations made by ASAS-SN. We wish to extend our special thanks to those of Hawaiian ancestry on whose sacred mountains of Maunakea and Haleakal\=a, we are privileged to be guests. Without their generous hospitality, the observations presented herein would not have been possible.\\

\textit{Facilities:} Laser Interferometer Gravitational-Wave Observatory (USA), Virgo (Italy), Haleakala Observatories (USA), Cerro Tololo International Observatory (Chile), McDonald Observatory (USA), South African Astrophysical Observatory (South Africa).\\

\textit{Software:} astropy \citep{astropy}, HEALPix \citep{healpix}, Healpy \citep{healpy}, Matplotlib \citep{matplotlib}, Numpy \citep{numpy}, Scipy \citep{scipy}

\section*{Data Availability Statements}
All the BAYESTAR maps used in this work to produce the figures are public and available on the Gravitational-Wave Candidate Event Database (GraceDB) which is a service operated by the LIGO Scientific Collaboration (\url{https://gracedb.ligo.org/superevents/public/O3/}).



\bsp	
\label{lastpage}
\end{document}